\RequirePackage{fix-cm}
\documentclass[smallextended]{svjour3}       % onecolumn (second format)
\smartqed  % flush right qed marks, e.g. at end of proof
\usepackage{graphicx}
\begin{document}
\def\vc#1{\mbox{\boldmath $#1$}}

\title{Electric dipole moment of light nuclei
%\thanks{Grants or other notes
%about the article that should go on the front page should be
%placed here. General acknowledgments should be placed at the end of the article.}
}
\subtitle{
%Do you have a subtitle?\\ If so, write it here
$-$ $^6$Li, $^7$Li, $^9$Be, $^{11}$B, and $^{13}$C $-$
}

%\titlerunning{Short form of title}        % if too long for running head

\author{Nodoka Yamanaka
%         \and
%        Second Author %etc.
}

%\authorrunning{Short form of author list} % if too long for running head

\institute{N. Yamanaka \at
IPNO, CNRS-IN2P3, Univ. Paris-Sud, Universit\'e Paris-Saclay, 
91406 Orsay Cedex, France\\
%              first address \\
%              Tel.: +123-45-678910\\
%              Fax: +123-45-678910\\
              \email{yamanaka@ipno.in2p3.fr}           %  \\
%             \emph{Present address:} of F. Author  %  if needed
%           \and
%           S. Author \at
%              second address
}

\date{Received: date / Accepted: date}
% The correct dates will be entered by the editor

\maketitle

\begin{abstract}
In this proceeding, we present the results of the theoretical evaluations of the electric dipole moment (EDM) of light nuclei, including the preliminary value for the $^{11}$B nucleus.
From the data, we can infer an approximate counting rule, and predict the EDM of other light nuclei.
% \PACS{PACS code1 \and PACS code2 \and more}
% \subclass{MSC code1 \and MSC code2 \and more}
\end{abstract}

\section{Introduction}
\label{intro}

To generate the baryon number asymmetry, Sakharov's criteria \cite{sakharov} have to be fulfilled.
One of them, the CP violation, is actually insufficient in the standard model (SM), and contributions from new physics beyond the SM is required.
As a good probe of CP violation beyond the SM, the electric dipole moment (EDM) \cite{engeledmreview,yamanakabook,atomicedmreview,chuppreview} is widely studied.
Here we discuss the EDM of light nuclei which is expected to be accurately measured using storage rings \cite{storage1,storage2,anastassopoulos,devriesedmreview,jedi,yamanakanuclearedmreview}.

Theoretically, the nuclear EDM presents many advantages.
The bare nucleus has no atomic electrons, so there is no suppression of hadronic CP violation due to Schiff's screening \cite{schiff}.
The nuclear EDM is also almost free from the CKM contribution \cite{yamanakasmedm1,yamanakasmedm2}, so that it has very small SM backgrounds to be considered.

An interesting question is whether the sensitivity of the nuclear EDM on CP violation beyond the SM is enhanced by many-body effects.
Here we discuss the EDM of light nuclei, which can be treated in the cluster model \cite{clusterreview3} with good accuracy.
By analyzing the EDM of several light nuclei ($^2$H, $^3$He, $^6$Li, $^7$Li, $^9$Be, $^{11}$B, and $^{13}$C), we will try to deduce an approximate counting rule.
The preliminary result of the EDM of $^{11}$B is shown for the first time.

This proceeding is organized as follows.
In the next section, we show the setup of the cluster model.
In Section \ref{sec:nuclearedm}, we summarize the current results on the EDM of light nuclei, and deduce an approximate counting rule.
Section \ref{sec:summary} is devoted to the summary.

\section{The cluster model and the CP-odd nuclear force}

Let us first introduce the CP-even interaction used in the cluster model.
In our cluster model, the nucleons, the $\alpha$ ($^4$He) and triton ($^3$H, denoted as $t$) clusters are the relevant degrees of freedom.
For the $N-N$ interaction required for the $^6$Li nucleus, the A$v$8' potential \cite{av18} is used.
For the CP-even $\alpha -N$ and $\alpha -\alpha$ interactions, we use the Kanada-Kaneko potential \cite{kanada} and the modified Hasegawa-Nagata potential \cite{hasegawa}, respectively, which were obtained by fitting the data of low energy scattering experiments.
For the calculation of $^{13}$C, we use the Kanada-Kaneko and Schmid-Wildermuth \cite{schmid} potentials, augmented by phenomenological three- and four-cluster interactions to reproduce the binding energies of the ground state as well as those of subsystems \cite{yamada13c}.
We use the interaction of Nishioka {\it et al}. \cite{nishioka} for the CP-even $\alpha -t$ potential, required in the calculation of $^7$Li and $^{11}$B.
For the $^{11}$B nucleus, we also introduce a phenomenological $\alpha -\alpha -t$ interaction to reproduce the energy levels of the ground state ($3/2_1^-$) and the $1/2_1^+$ states \cite{yamada11b}.
In our cluster model calculations, the orthogonality condition model \cite{saito1,saito2,saito3} is applied to exclude forbidden states.

\begin{figure}
\includegraphics[width=0.75\textwidth]{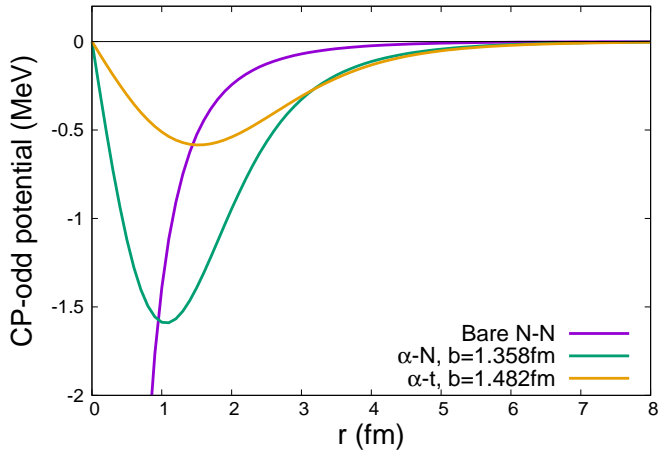}
\caption{
Radial dependence of the CP-odd nuclear interactions.
}
\label{fig:folding}
\end{figure}

Let us now model the CP-odd interaction.
For the bare $N-N$ system, the leading CP-odd Hamiltonian is given by the following one-pion exchange potential with three possible isospin structures \cite{pvcpvhamiltonian3}:
\begin{eqnarray}
H_{P\hspace{-.45em}/\, T\hspace{-.5em}/\, }^\pi
& = &
\bigg\{ 
\bar{G}_{\pi}^{(0)}\,{\vc{\tau}}_{1}\cdot {\vc{\tau}}_{2}\, {\vc{\sigma}}_{-}
+\frac{1}{2} \bar{G}_{\pi}^{(1)}\,
( \tau_{+}^{z}\, {\vc{\sigma}}_{-} +\tau_{-}^{z}\,{\vc{\sigma}}_{+} )
\nonumber\\
&&\hspace{6em}
+\bar{G}_{\pi}^{(2)}\, (3\tau_{1}^{z}\tau_{2}^{z}- {\vc{\tau}}_{1}\cdot {\vc{\tau}}_{2})\,{\vc{\sigma}}_{-} 
\bigg\}
\cdot
\frac{ \vc{r}}{r} \,
V(r)
,
\label{eq:CPVhamiltonian}
\end{eqnarray}
where $\vc{r} \equiv \vc{r}_1 - \vc{r}_2$ is the relative coordinate.
The spin and isospin matrices are defined by ${\vc{\sigma}}_{-} \equiv {\vc{\sigma}}_1 -{\vc{\sigma}}_2$, ${\vc{\sigma}}_{+} \equiv {\vc{\sigma}}_1 + {\vc{\sigma}}_2$, ${\vc{\tau}}_{-} \equiv {\vc{\tau}}_1 -{\vc{\tau}}_2$, and ${\vc{\tau}}_{+} \equiv {\vc{\tau}}_1 + {\vc{\tau}}_2$.
The radial behavior of the CP-odd potential is given by
\begin{equation}
V (r)
= 
-\frac{m_\pi}{8\pi m_N} \frac{e^{-m_\pi r }}{r} \left( 1+ \frac{1}{m_\pi r} \right)
\ .
\end{equation}
Its shape is shown in Fig. \ref{fig:folding}.

In the point-of-view of the chiral perturbation theory, the CP-odd potential of Eq. (\ref{eq:CPVhamiltonian}) is not complete.
In the leading order, there are also P and CP violating contact CP-odd $N-N$ interactions \cite{cpveft} as well as a CP-odd three-nucleon interaction \cite{cpveft2}.
The contribution of the former to the nuclear EDM, however, has a large uncertainty due to the poorly known short-range physics of nuclei \cite{bsaisou}, so we do not consider it.
The three-nucleon interaction is also neglected due to its small contribution to the EDM of light nuclei \cite{bsaisou}.
This is due to the fact that light nuclei have a dominant configuration with paired nucleons, whereas the CP-odd three-nucleon force requires three nucleons with aligned spins.
We also note that the isotensor CP-odd nuclear force [term with $\bar{G}_{\pi}^{(2)}$ in Eq. (\ref{eq:CPVhamiltonian})] is subleading in chiral perturbation, but we display it by habit.

The CP-odd $\alpha -N $ and $\alpha -t$ potentials are modeled by folding \cite{horiuchi} the CP-odd $N-N$ interaction (\ref{eq:CPVhamiltonian}).
The isoscalar and isotensor CP-odd interactions cancel due to the closure of the spin and isospin shells of the $\alpha$-clusters.
For the oscillator parameters, we take $b = 1.358$ fm ($\alpha - N$) and $b=1.482$ fm ($\alpha - t$).
The shape of the folding CP-odd potentials are displayed in Fig. \ref{fig:folding}.

\section{The nuclear electric dipole moment: definition and results}
\label{sec:nuclearedm}

The nuclear EDM is generated by two leading contributions.
The first one is given by the intrinsic EDM of the nucleon:
\begin{eqnarray}
d_{A}^{\rm (Nedm)} 
&=&
\sum_{i=1}^{A} \frac{1}{2} 
\langle \, \Phi_J (A) \, |\, (1+\tau_i^z ) \, \sigma_{iz} \, | \, \Phi_J (A) \, \rangle
,
\label{eq:polarizationedm}
\end{eqnarray}
where $|\, \Phi_{J} (A)\, \rangle$ is the state vector of the polarized nucleus $A$.
The second process is the polarization of the nucleus by the CP-odd nuclear force:
\begin{eqnarray}
d_{A}^{\rm (pol)} 
&=&
\sum_{i=1}^{A} \frac{e}{2} 
\langle \, \Phi_J (A) \, |\, (1+\tau_i^z ) \, r_{iz} \, | \, \Phi_J (A) \, \rangle
.
\label{eq:polarizationedm}
\end{eqnarray}
Since the nucleon EDM and the CP-odd nuclear force are very small, the nuclear EDM can be expressed in the leading order of perturbation as
\begin{eqnarray}
d_{A} 
&=&
\langle \sigma_p \rangle d_p
+\langle \sigma_n \rangle d_n
+\bar G_\pi^{(0)}
a_\pi^{(0)}
+\bar G_\pi^{(1)}
a_\pi^{(1)}
+\bar G_\pi^{(2)}
a_\pi^{(2)}
.
\label{eq:polarizationedm}
\end{eqnarray}
Evidently, the most interesting nuclei are those which have large coefficients.

\begin{table}
\caption{
The linear coefficients of the nuclear EDM for several nuclei.
The symbol ``$-$'' means that either the coefficient cancels or it cannot be determined with sufficient accuracy.
}
\label{tab:nuclearedm}
\begin{center}
\begin{tabular}{llllll}
\hline\noalign{\smallskip}
 & $\langle \sigma_p \rangle$ & $\langle \sigma_n \rangle$ &$a_\pi^{(0)}$ ($ e$ fm) & $a_\pi^{(1)}$ ($e$ fm) & $a_\pi^{(2)}$ ($ e$ fm) \\ 
\noalign{\smallskip}\hline\noalign{\smallskip}
$^{2}$H \cite{liu,yamanakanuclearedm} & 0.91 & 0.91 & $-$ & $0.0145 $ & $-$  \\
$^{3}$He \cite{bsaisou,yamanakanuclearedm}& $-0.04$ & 0.89 & $0.0059$ & 0.0108 & 0.0168  \\
$^{3}$H \cite{bsaisou,yamanakanuclearedm} & 0.88 & $-0.05$ & $-0.0059$ & 0.0108 & $-0.0170$  \\
$^{6}$Li \cite{yamanakanuclearedm}& 0.86 & 0.86 & $-$ & 0.022 & $-$  \\
$^{7}$Li \cite{PPNS} & 0.9 & $-$ & $-0.006$ & 0.016 & $-0.017$  \\
$^{9}$Be \cite{yamanakanuclearedm}  & $-$ & 0.75 & $-$ & $0.014$ & $-$  \\
$^{11}$B & 0.7 & $-$ & $-0.004$ & $0.02$ & $-0.01$  \\
$^{13}$C \cite{c13edm} & $-$ & $-0.33$ & $-$ & $-0.0020 $ & $-$  \\
$^{129}$Xe \cite{yoshinaga1,yoshinaga2}  & $-$& 0.2 & $7 \times 10^{-5}$ & $7 \times 10^{-5} $ & $4 \times 10^{-4}$  \\
\noalign{\smallskip}\hline
\end{tabular}
\end{center}
\end{table}

\begin{figure}
\includegraphics[width=1.0\textwidth]{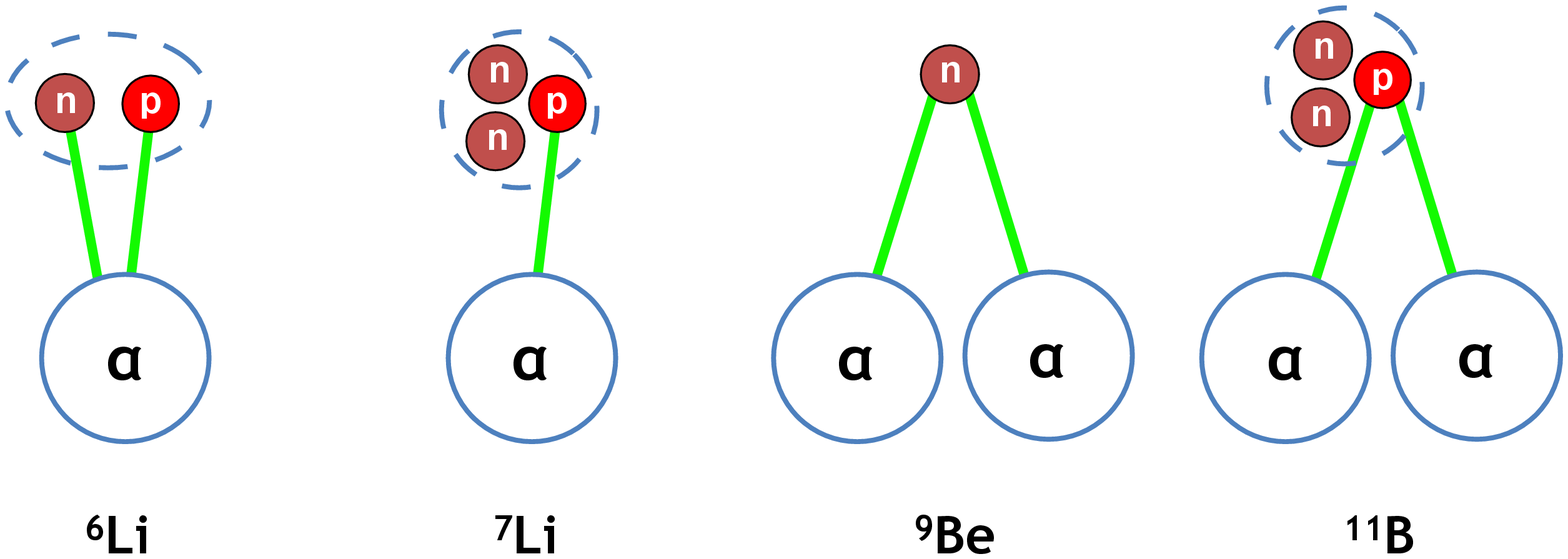}
\caption{
Counting rule for the EDM of light nuclei.
The EDM is induced by the constructive interference between the EDM of the deuteron or triton clusters and the CP-odd $\alpha -N$ or $\alpha -t$ polarizations.
}
\label{fig:counting_rule}
\end{figure}

In Table \ref{tab:nuclearedm}, we list the coefficients of the EDMs of $^6$Li \cite{yamanakanuclearedm}, $^7$Li \cite{PPNS}, $^9$Be \cite{yamanakanuclearedm}, $^{11}$B, and $^{13}$C \cite{c13edm}, calculated in the cluster model.
The result of $^{11}$B is new and preliminary.
Those of the deuteron \cite{liu,yamanakanuclearedm} and three-nucleon systems \cite{bsaisou,yamanakanuclearedm}, calculated with the phenomenological nuclear force Argonne $v18$ \cite{av18}, have also been displayed for comparison.
We see that several nuclei, such as the $^6$Li, $^7$Li, or $^{11}$B, have larger isovector coefficients than that of the deuteron.
This enhancement can be understood as the constructive interference between the EDM of the deuteron or triton cluster and the polarization from the CP-odd $\alpha -N$ or $\alpha -t$ interactions (see Fig. \ref{fig:counting_rule}).
We determine the latter by equating the following system of equations:
\begin{eqnarray}
d_{^{6}{\rm Li}}
& = &
2\times (\alpha -N \, \mbox{polarization})
+ d_{^{2}{\rm H}}
,
\nonumber\\
d_{^{7}{\rm Li}}
& = &
1 \times (\alpha -N \, \mbox{polarization})
+d_{^{3}{\rm H}}
,
\nonumber\\
d_{^{9}{\rm Be}}
& = &
2 \times (\alpha -N \, \mbox{polarization})
,
\nonumber\\
d_{^{11}{\rm B}}
& = &
2 \times (\alpha -N \, \mbox{polarization})
+d_{^{3}{\rm H}}
.
\end{eqnarray}
After using the values of Table \ref{tab:nuclearedm}, one obtains $(\alpha -N \, \mbox{polarization}) \sim (0.005 - 0.007)\, \bar G_\pi^{(1)}e$ fm.
This result forms an approximate counting rule for the EDM of light nuclei.
From this, we can predict the EDM of heavier nuclei.
For instance, we have
\begin{eqnarray}
d_{^{10}{\rm B}}
& \sim &
4\times (\alpha -N \, \mbox{polarization})
+ d_{^{2}{\rm H}}
\sim
0.03 \, \bar G_\pi^{(1)}e \, {\rm fm}
,
\nonumber\\
d_{^{14}{\rm N}}
& \sim &
6 \times (\alpha -N \, \mbox{polarization})
+ d_{^{2}{\rm H}}
\sim
0.04 \, \bar G_\pi^{(1)}e \, {\rm fm}
.
\end{eqnarray}

The contribution from the EDM of the deuteron cluster in $^6$Li and that of the triton in $^{11}$B are slightly decreased due to the mixing of angular momentum configurations.
The dependence on the isoscalar and isotensor nuclear forces of the EDM of $^7$Li and $^{11}$B is due to the triton cluster.
It may additionally be given by the intrinsic EDM of the nucleon, although we do not discuss it in this work.
It is also important to note that the two spins of the neutrons in the triton are likely to form a singlet, so they do not contribute to the CP-odd $\alpha - t$ polarization.

For the EDM of $^{13}$C, however, we observe that the isovector coefficient is small \cite{c13edm}, and the counting rule cannot be applied.
The $^{13}$C nucleus has a ground state ($1/2^-_1$) composed by a dominant configuration of a $^{12}$C core with angular momentum two.
It happens that this state cannot easily make transition with the closest opposite parity state $1/2^+_1$ which is 3.1 MeV above it through CP-odd operators \cite{c13edm}.
The ground state is rather coupled with a state which is separated by about 10 MeV, which suppresses the EDM.
This feature is also expected to be relevant for $^{15}$N which also has a similar level structure.

What will happen when we increase the nucleon number?
Na\"{i}vely, we can imagine that the contribution from the CP-odd $\alpha -N$ polarization will grow.
This will however not be realized in real nuclei, since the configuration mixing, which makes destructive interference between angular momentum configurations of valence nucleons, is relevant in heavy nuclei.
In Table \ref{tab:nuclearedm}, we see that the result for the nuclear EDM of $^{129}$Xe is much smaller than those of light nuclei \cite{yoshinaga1,yoshinaga2}.
Moreover, heavy nuclei are more difficult to handle in storage ring experiments, so they have no remarkable advantages.

\section{Summary}
\label{sec:summary}

In this proceeding, we presented the calculations of the EDM of light nuclei.
The preliminary result of the EDM of $^{11}$B was shown for the first time.
By equating the sensitivity of the EDM of $^{6}$Li, $^{7}$Li, $^{9}$Be, and $^{11}$B on the isovector CP-odd nuclear force, we could infer an approximate counting rule.
We could then estimate the EDM of other unknown nuclei such as $^{10}$B or $^{14}$N which are estimated to be more sensitive than known ones.

On the other hand, there are other nuclei such as $^{13}$C for which the sensitivity on the isovector CP violation is suppressed by their nuclear structure.
Finally, increasing the number of nucleons is not likely to provide us sensitive nuclei on CP violation, since the configuration mixing will suppress the polarization.
We conclude for the moment that light nuclei are the most suitable to be measured in storage ring experiments.

\begin{acknowledgements}
The author is supported by the JSPS Postdoctoral Fellowships for Research Abroad.
\end{acknowledgements}

% BibTeX users please use one of
%\bibliographystyle{spbasic}      % basic style, author-year citations
%\bibliographystyle{spmpsci}      % mathematics and physical sciences
%\bibliographystyle{spphys}       % APS-like style for physics
%\bibliography{}   % name your BibTeX data base

% Non-BibTeX users please use

\end{document}